\documentclass[10pt,conference]{IEEEtran}
\IEEEoverridecommandlockouts
\usepackage{cite}
\usepackage{amsmath,amssymb,amsfonts}
\usepackage{algorithmic}
\usepackage{graphicx}
\usepackage{textcomp}
\usepackage{xcolor}
\def\BibTeX{{\rm B\kern-.05em{\sc i\kern-.025em b}\kern-.08em
    T\kern-.1667em\lower.7ex\hbox{E}\kern-.125emX}}

\usepackage{enumitem}
\usepackage{balance}
\usepackage{tcolorbox}
\usepackage{microtype}
\usepackage{multirow}
\usepackage{booktabs}
\usepackage{flushend}
\usepackage{listings}

\newcommand{\ind}[1]{\vspace{5pt}\noindent\textbf{#1}}

\usepackage{xcolor}
\usepackage{url}

\definecolor{gray50}{gray}{.5}
\definecolor{gray40}{gray}{.6}
\definecolor{gray30}{gray}{.7}
\definecolor{gray20}{gray}{.8}
\definecolor{gray10}{gray}{.9}
\definecolor{gray05}{gray}{.95}

\newlength\Linewidth
\def\findlength{\setlength\Linewidth\linewidth
\addtolength\Linewidth{-4\fboxrule}
\addtolength\Linewidth{-3\fboxsep}
}

\newenvironment{examplebox}{\par\begingroup
  \setlength{\fboxsep}{5pt}\findlength
  \setbox0=\vbox\bgroup\noindent
  \hsize=0.94\linewidth
  \begin{minipage}{0.94\linewidth}\normalsize}
    {\end{minipage}\egroup
    \textcolor{gray20}{\fboxsep1.5pt\fbox
     {\fboxsep5pt\colorbox{gray05}{\normalcolor\box0}}}
    \endgroup\par\noindent
    \normalcolor\ignorespacesafterend}

\usepackage{framed}
\newcommand{\answer}[2]{%
\begin{framed}%
\noindent \textbf{Answer to RQ{#1}: }{#2}%
\end{framed}
}

\usepackage{framed}
\newcommand{\finding}[2]{%
\begin{framed}%
\noindent \textbf{Finding {#1}: }\textit{{#2}}%
\end{framed}
}

\usepackage{color}

\definecolor{mygreen}{rgb}{0,0.6,0}
\definecolor{mygray}{rgb}{0.5,0.5,0.5}
\definecolor{mymauve}{rgb}{0.58,0,0.82}

\makeatletter
\AtBeginDocument{\let\c@lstlisting\c@figure}
\makeatother

\begin{document}

\title{Improving ChatGPT Prompt for Code Generation}



\author{
\IEEEauthorblockN{Chao Liu$^1$, Xuanlin Bao$^1$, Hongyu Zhang$^1$, Neng Zhang$^2$, Haibo Hu$^1$, Xiaohong Zhang$^1$, Meng Yan$^1$}
\IEEEauthorblockA{\textit{$^1$School of Big Data and Software engineering, Chongqing University, China} \\
\textit{\{liu.chao, baoxuanlin, hyzhang, haibo.hu, xhongz, mengy\}@cqu.edu.cn}\\
\textit{$^2$School of Software Engineering, Sun Yat-sen University, China}\\
\textit{zhangn279@mail.sysu.edu.cn}\\}
}
\newcommand{\hy}[1]{\textcolor{red}{HY:{#1}}}
\pagestyle{plain}

\maketitle

\begin{abstract}

Automated code generation can be a powerful technique for software development, significantly reducing developers' efforts and time required to create new code by generating it automatically based on requirements. Recently, OpenAI's language model ChatGPT has emerged as a powerful tool for generating human-like responses to a wide range of textual inputs (i.e., prompts), including those related to code generation. However, the effectiveness of ChatGPT for code generation is not well understood, and the generation performance could be heavily influenced  by the choice of prompt. To answer these questions, we conducted experiments using the CodeXGlue dataset to evaluate ChatGPT's capabilities for two code generation tasks, including text-to-code and code-to-code generation. We designed prompts by leveraging the chain-of-thought strategy with multi-step optimizations. Our results showed that by carefully designing prompts to guide ChatGPT, the generation performance can be improved substantially. We also analyzed the factors that influenced the prompt design and provided insights that could guide future research.

\end{abstract}

\begin{IEEEkeywords}
ChatGPT, code generation, prompt engineering
\end{IEEEkeywords}

\section{Introduction}\label{intro}
Code generation is a technique that aims to automatically generate code based on developers' requirements \cite{herrington2003code,poesia2022synchromesh}. It can reduce repetitive coding efforts and improve software development productivity \cite{xu2020incorporating,guo2020graphcodebert}. These requirements can be expressed as natural language (NL) descriptions, allowing developers to specify their needs in an intuitive way. For instance, a developer can ask a code generation tool to \textit{"convert an integer variable n to a string in Java"}, and the tool will generate an appropriate code example such as: \textit{"String s = Integer.toString(n)"}. This process is known as Text-to-Code (T2C) generation \cite{lu2021codexglue,nijkamp2022codegen}. Another type of code generation is Code-to-Code (C2C) generation, which translates an existing code snippet from one programming language to another \cite{liu2022commitbart,lu2021codexglue}. For instance, the C\# code \textit{"String s = n.ToString()"} can be translated to the above Java code. The C2C generation can be useful when porting existing code to a new programming language \cite{liu2022commitbart}.



Large language models (LLMs) have emerged as a powerful tool for natural language processing (NLP) tasks, such as sentiment analysis \cite{araci2019finbert,zhou2020sentix} and language translation \cite{gunel2020supervised}, thanks to their ability to be pre-trained on massive amounts of massive unsupervised textual data and fine-tuned on domain-specific datasets. This \textit{"pre-train, fine-tune"} paradigm has been applied to software engineering (SE) tasks, such as code generation, with promising results. For instance, Feng et al. \cite{feng2020codebert} developed CodeBERT, an LLM that has a similar architecture to BERT \cite{devlin2018bert} and is pre-trained on six programming languages \cite{devlin2018bert}. CodeBERT can be used for various SE tasks, such as code search and summarization, with good performance. Another notable model is CodeGPT developed by Lu et al. \cite{lu2021codexglue}. CodeGPT is pre-trained on Python and Java datasets using the GPT-2 \cite{raffel2020exploring} architecture, and fine-tuned for a variety of SE tasks, such as code generation and code translation. 



Recently, OpenAI introduced ChatGPT, a revolutionary LLM based on the GPT-3.5 architecture \cite{openai2023chatgpt}, which can work on various tasks including code generation \cite{khoury2023secure}. Different from existing LLMs, ChatGPT is able to generate human-like responses through reinforcement learning \cite{ouyang2022training} based on users' textual inputs (i.e., prompts). Owing to its effectiveness on various tasks, ChatGPT has attracted 100 million active users worldwide within just two months after its initial release \cite{kothari2023chatgpt}. However, the performance of ChatGPT is highly dependent on the quality of prompts used. Designing better prompts, which is called prompt engineering \cite{liu2022design}, is under active investigation. In this paper, we investigate the code generation performance of ChatGPT with various prompt engineering methods. 


We conducted an evaluation of ChatGPT's code generation capabilities using the widely used CodeXGlue \cite{lu2021codexglue} dataset for both T2C and C2C generation tasks. Initially, we employed basic prompts for the tasks: \textit{"write a Java method that" + NL description} for T2C task, and \textit{"translated C\# code into Java code:" + code} for C2C task. Experimental results showed that these prompts (ChatGPT-task) achieved CodeBLEU scores of 22.76 and 39.37, respectively, where CodeBLEU is a widely used overall evaluation metric \cite{ren2020codebleu}. 
To improve the generation performance, we leveraged the chain-of-thought strategy with manual construction \cite{wei2022chain} to augment the prompts for different tasks. This approach conducts multi-step optimizations based on the feedback from ChatGPT. Our experimental results showed that: 1) adding more specific requirements to the prompts improved the CodeBLEU of ChatGPT-task by 73.58\% and 3.45\% for the two tasks, respectively; 2) directly asking ChatGPT to generate concise code in the prompt (e.g., \textit{"write a concise Java method that" + NL}) led to further improvement in CodeBLEU for the T2C task, reaching to 50.18; 3) sharing a ChatGPT session for a number of prompt testing also boosted the CodeBLEU of the C2C task to 48.80; and 4) the generation randomness of ChatGPT had little effect on the generation performance due to the specific instructions in the prompt. Furthermore, we compared the performance with state-of-the-art fine-tuned LLMs and analyzed the correctness and quality of the generated code. 


In summary, the major contributions of this paper are as follows:

\begin{itemize}
    \item Evaluating ChatGPT on a widely-used dataset CodeXGlue for two code generation tasks.\vspace{3pt}

    \item Proposing prompt design and optimization methods to 
    guide ChatGPT to generate better code with prompt engineering.\vspace{3pt}

    \item Releasing a replication package\footnote{Replication Package: https://anonymous.4open.science/r/guiding-chatgpt-for-code-generation-0B0E}  for future exploration in this research community.
\end{itemize}


\section{Background and Related Work}\label{related}

\subsection{Large Language Model for Code Generation}\label{sec_llm}



Many language models (LMs) have been proposed, which are pre-trained with a special objective (e.g., masked language modeling \cite{salazar2019masked}) and applied to downstream tasks by fine-tuning.  Generally, there are three types of LMs: \textit{1) Masked LM,} a model is trained to predict a masked word in a sentence given its surrounding contexts, such as BERT \cite{devlin2018bert} and RoBERTa \cite{liu2019roberta}. \textit{2) Encoder-Decoder,} a model works for sentence-to-sentence takes like translation and summarization, where an encoder encodes the input into a fixed-length vector and a decoder generates output from the encoded vector, such as T5 \cite{raffel2020exploring}, BART \cite{alokla2022pseudocode}, and MASS \cite{niu2022spt}. \textit{3) Left-to-Right LM,} a model is trained to predict the next word in a sentence given the previous words, such as GPT \cite{radford2019language}, GPT-2 \cite{raffel2020exploring}, and GPT-3 \cite{brown2020language}. For these LMs, Transformer \cite{vaswani2017attention} is used as the base model because its self-attention layers can efficiently process input with long-term memory and effectively adapt itself to various downstream tasks \cite{liu2023pre}.

Researchers have proposed many LM-based models that can be used for code generation tasks. The representatives are: \textit{1) BERT-Based.} CodeBERT \cite{feng2020codebert} trained a BERT-like model with six programming languages. GraphCodeBERT \cite{guo2020graphcodebert} is an improved model that considers the inherent structure of code instead of plain text as CodeBERT. UniXcoder \cite{guo2022unixcoder} addressed the difficulty in learning code structure by transforming the code into a sequence but retaining the structural information. ContraBERT \cite{liu2023contrabert} leveraged contrastive learning \cite{jain2020contrastive} to improve the robustness of CodeBERT and GraphCodeBERT. \textit{2) T5-Based.} Mastropaolo et al. \cite{mastropaolo2021studying} showed that fine-tuning T5 is possible to work on SE tasks. CodeT5 \cite{wang2021codet5} is an identifier-aware T5 model that can distinguish which code tokens are identifiers and recover them when they are masked. \textit{3) BART-Based.} PLBART \cite{ahmad2021unified} is constructed on BART pre-trained with an extensive collection of Java and Python functions and associated NL text via denoising autoencoding. CommitBART \cite{liu2022commitbart} pre-trained BART using data collected from GitHub commits. \textit{4) GPT-Based.} GPT-C \cite{svyatkovskiy2020intellicode} is a variant of GPT-2 pre-trained on a large unsupervised multilingual source code dataset. CodeGPT \cite{lu2021codexglue} pre-trained GPT-2 on Python and Java corpora from the CodeSearchNet \cite{husain2019codesearchnet}. CodeGen \cite{nijkamp2022codegen} presents a family of architectures similar to GPT-3 designed for multi-turn program synthesis. CodeX \cite{chen2021evaluating} fine-tuned GPT-3 on publicly available code from GitHub, whose distinct production version powers the GitHub Copilot \cite{github2023copilot}. 

\subsection{ChatGPT and Prompt Engineering}\label{sec_pe}
ChatGPT is an LM developed by OpenAI and it is designed for conversational tasks (e.g., question-answering and code generation) \cite{openai2023chatgpt}. ChatGPT is built on the GPT-3.5 series with 175 billion parameters and optimized by using reinforcement learning from human feedback \cite{ouyang2022training}. It can generate human-like responses to the user's textual prompt based on its context understanding and conversation history. Besides, OpenAI is improving ChatGPT by keeping optimizing GPT-4 \cite{openai2023gpt4}.

As LM (e.g., ChatGPT \cite{openai2023chatgpt}) with a large number of parameters ($>$100 million) emerges with advanced textual generation capability, prompt engineering (PE) becomes a new paradigm for NLP \cite{liu2023pre}. The goal of PE is to design an appropriate prompt for a pre-trained model and conduct prediction as expected with good performance, leading to the \textit{"pre-train, prompt, predict"} paradigm. Specifically, the PE creates a prompt $x'=f_{prompt}(x)\in{X}$ for a textual input $x$ (e.g., "write a Java code for converting int to string") that describes a downstream task (e.g., code generation). With the prompt $x$, LLM performs prediction $y=f_{LLM}(x')\in{Y}$. Two basic PE tasks are: \textit{1) Prompt Template Engineering,} it designs an appropriate template $x'$ for the LM input (e.g., "write a Java code for [x]", where "[x]" is a variable for NL description), as the performance of LM prediction $y$ is sensitive to sentence(s) designed in the template. \textit{2) Prompt Answer Engineering,} it aims to design an answer space $Z$ in the prompt so that a better answer $y$ could be generated from a limited scope $y\in{Z}$ (e.g., "Which code is better? A or B"). More advanced PE tasks intend to manipulate multi-prompt, such as prompt augmentation \cite{rajagopal2021template}, composition, etc. \cite{liu2023pre}

The prompt ($x'$) can be generated in four ways ($f_{prompt}$) \cite{qiao2022reasoning,liu2023pre}: \textit{1) Manual Construction,} it is suitable for template-based prompts and few-shot prompting where the prompt is uncomplicated \cite{rajagopal2021template,wei2022emergent}. \textit{2) LM Generation,} it leverages LM to generate customized prompt ($x'$) for each textual input ($x$), which can make up for the shortcomings of the manual construction \cite{zhang2022automatic}. \textit{3) Retrieval-Based Prompt,} it relies on well-annotated external resources (e.g., Wikipedia) to alleviate the unstable issue of generation \cite{mishra2022numglue}. \textit{4) Prompt Learning,} it builds a supervised model to automatically update the prompt according to the LM's generation and the associated ground-truth \cite{liu2023pre}. In this study, we leveraged the manual construction to explore the possibility to guide ChatGPT for code generation tasks, investigate the influential factors in prompt design, and provide researchers with insights for future works.

\section{Study Subjects}\label{study}
To apply ChatGPT to code generation, this section presents the study subjects. Specifically, Section \ref{sec_task} describes two investigated code generation tasks. Section \ref{sec_dataset} presents the used dataset and Section \ref{sec_metrics} lists the evaluation metrics.

\subsection{Code Generation Tasks}\label{sec_task}
Code generation is the process of automatically generating code according to a requirement specification. Code generation can save time and efforts for developers by automating repetitive programming tasks. The specification can be expressed in different ways. In this study, we investigated two representative tasks: \textit{1) Text-to-Code (T2C) Generation.} It takes a textual description written in natural language (NL) as a functional specification. A code generation model generates code (e.g., Java) according to the textual description. \textit{2) Code-to-Code (C2C) Generation.} It takes a code snippet (e.g., C\#) as input and an NL model generates code written in another programming language (e.g., Java) with the same functionality. This task is also called a code translation due to its similarity to language translation. In the following two subsections, we will introduce the related models.

\subsection{Datasets}\label{sec_dataset}
Here we present the datasets for testing the investigated two types of code generation tasks.

\ind{T2C Dataset.} We chose the widely used dataset CONCODE \cite{iyer2018mapping}, which is collected in CodeXGLUE \cite{lu2021codexglue}. This dataset collected about 33k Java repositories from GitHub, consisting of 100k training data, 2k valid data, and 2k test data. Each instance in the data is a tuple of three elements: \textit{1) Code Snippet}, it is the ground-truth for code generation; \textit{2) Natural Language Description,} it is extracted from the Javadoc of the code snippet for generation input; \textit{3) Code Environment,} it describes the class file (i.e., the programmatic context) where the code snippet works, including class name, class path, member variables, and signatures of member functions.

\ind{C2C Dataset.} We used the C2C dataset from CodeXGLUE \cite{lu2021codexglue}, a popular benchmark dataset for code understanding and generation. The C2C dataset collected data from several open-source repositories, including Lucene, POI, JGit, and Antlr. Totally, it contains 10k training data, 0.5k valid data, and 1k test data. Each instance in a data contains a pair of code snippets written in C\# and Java, which shared the same function but are implemented in different programming languages. In this study, we regarded the Java code snippet as the generation target of T2C dataset, and used the C\# code snippet as the input.

\subsection{Evaluation Metrics}\label{sec_metrics}

To analyze the effectiveness of code generation, we measured the performance with the widely used metrics for code generation task \cite{lu2021codexglue}, including BLEU \cite{papineni2002bleu} and CodeBLEU \cite{ren2020codebleu}. Details are described as follows. Following Lu et al. \cite{lu2021codexglue}, we used the CodeBLEU as the overall evaluation metric.


\ind{BLEU,} a popular metric to measure the generation accuracy for the code snippets with various lengths \cite{lu2021codexglue,papineni2002bleu}. Specifically, $BLEU=BP*e^{(logP_{1}+...+logP_{n})/n}$ where $BP$ is the brevity penalty value, which equals 1 if the generated code is longer than the ground-truth. Otherwise, it equals the ratio between the lengths of two code. $P_{i}$ is the metric for the overlapping between the bags of $i$-grams  appearing in the generated code and the ground-truth. 

\ind{CodeBLEU,} a variant of BLEU, which also considers the syntactic and semantic data flow correctness of code generation. It is similar to BLEU, but it calculates precision scores based on a set of code tokens, rather than natural language n-grams. Generally, CodeBLEU is a weighted average of the lexical, abstract syntax tree, and data flow match between the generated code and the ground-truth \cite{ren2020codebleu}.

\section{Methodology}\label{method}
This section describes prompt engineering for two code generation tasks. Specifically, Section \ref{sec_process} describes the general method for prompt design. Sections \ref{sec_design_t2c} and \ref{sec_design_c2c} elaborate on the specific prompt design for two tasks, respectively. Finally, Section \ref{sec_design_rqs} presents the investigated research questions (RQs) in this study.

\subsection{Methods for Prompt Design}\label{sec_process}

The performance of ChatGPT is often sensitive to the design of prompts \cite{liu2023pre}. To augment the prompt, Wei et al. \cite{wei2022emergent} indicated that Chain-of-Thought (CoT) prompting is the key strategy, which enables an LLM to solve problems by guiding them to produce a sequence of intermediate steps before giving the final answer. Due to its effectiveness, the CoT strategy is widely investigated and applied \cite{wei2022chain,suzgun2022challenging}. 

Generally, to guide ChatGPT for code generation tasks, we designed the prompt with the CoT strategy in two steps: \textit{1) Prompt Description,} we first analyze the requirement of a code generation task, and design a basic prompt in a natural way. Then, we provide the basic prompt for ChatGPT and ask \textit{"how to improve the prompt?"}, and further improve the prompt according to ChatGPT's suggestions. \textit{2) Multi-Step Optimizations,} we test the prompt in the first step on some samples from training data of the related dataset, analyze the generation performance with the ground-truth, and keep optimizing the generation results by providing ChatGPT with a series of new prompts. 

\begin{figure}
    \centering
    \includegraphics[width=\linewidth]{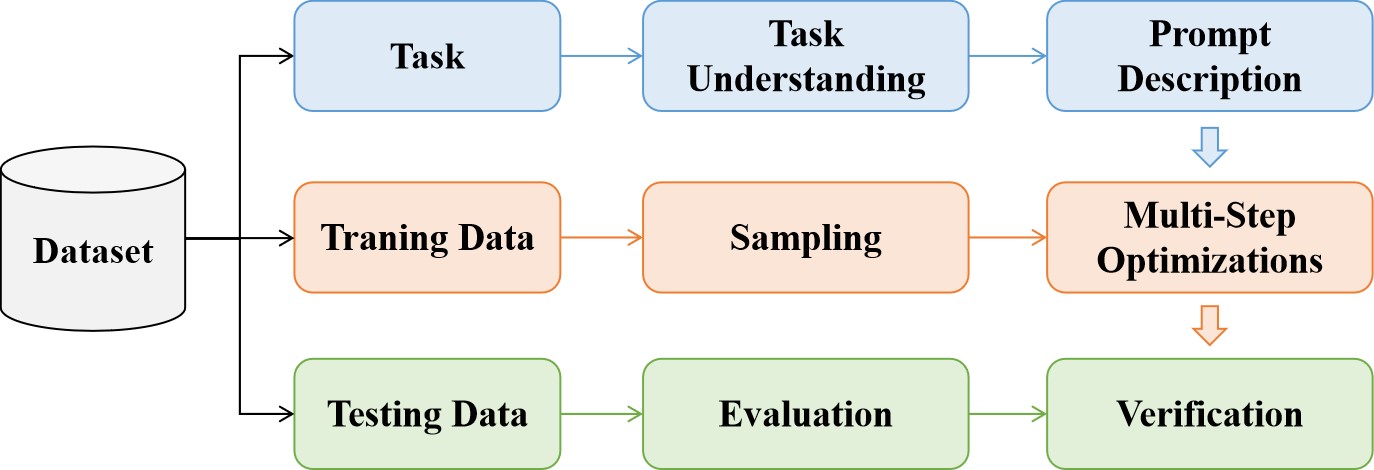}
    \caption{Overview of the method}
    \label{fig_overview}
\end{figure}

Based on the knowledge of the prompt design process, we generated some baseline prompts and evaluated them on the testing data, which can be found in Section \ref{rq1}. Fig. \ref{fig_overview} illustrates the overview of 
prompt design and verification. During the prompt design and testing, we work with ChatGPT by invoking its API \cite{openai2023api} with default settings (e.g., using the GPT-3.5-Turbo model). Table \ref{tab_prompt_sample} shows the generation performance of ChatGPT using different combinations of prompts designed in Table \ref{tab_prompt_type}. The following two sections elaborate on how we design prompts for two code generation tasks, respectively, where the discussed prompts are listed in Table \ref{tab_prompt_type}.

\begin{table*}
    \centering
    \caption{Different types of prompts designed for two code generation tasks. Note that \#\{NL\}, \#\{CN\}, \#\{MV\}, \#\{MF\}, and \#\{Code\} stand for the variables of a class name, member variable, member function, and code, which will be filled in actual inputs from the dataset.}
    \label{tab_prompt_type}
    \begin{tabular}{c|p{0.14\linewidth}|p{0.37\linewidth}|p{0.37\linewidth}}
        \toprule
        \textbf{No.} & \textbf{Prompt Type} & \textbf{Text-to-Code Generation Task} & \textbf{Code-to-Code Generation Task}\\
        \midrule
        P1&Task Prompt & write a Java method that + \#\{NL\} & translate C\# code into Java code: \#\{Code\} \\
        \midrule
        P2&Context Prompt & remember you have a Java class named + '\#\{CN\}', member variables + '\#\{MV\}', member functions + '\#\{MF\}' & - \\ 
        \midrule
        P3&Processing Prompt & remove comments; remove summary; remove throws; remove function modifiers; change method name to "function"; change argument names to "arg0", "arg1"...; change local variable names to "loc0", "loc1"... & do not provide annotation\\
        \midrule
        P4&Updated Task Prompt & - & translate C\# code delimited by triple backticks into Java code: '''\#\{Code\}''' \\
        \midrule
        P5&Behaviour Prompt & write a Java method \#\{that calls ...\} with[out] exception handling to \#\{NL\} & translate C\# code into Java code: '''\#\{Code\}''' \#\{that calls ...\} with[out] exception handling \\
        \bottomrule
    \end{tabular}
\end{table*}

\begin{table}[]
    \centering
    \caption{Testing different prompt combinations in Table \ref{tab_prompt_type} on 100 samples randomly selected from training data of each generation task. Note that P5(API) indicates that we only used the API part of the prompt P5.}
    \begin{tabular}{l|p{0.27\linewidth}|ll}
        \toprule
        \textbf{Task} & \textbf{Model} & \textbf{BLEU} & \textbf{CodeBLEU} \\
        \midrule
        \multirow{5}{*}{T2C}
        & P1 & 05.29 & 22.76\\
        & P2+P1 & 10.42 (+96.98\%)  & 25.05 (+10.06\%)  \\
        & P2+P1+P3 & 13.11 (+147.83\%) & 36.00 (+58.17\%)\\
        & P2+P5(API)+P3 & 22.14 (+318.53\%) & 44.18 (+94.11\%)\\
        & P2+P5+P3 & 27.48 (+419.47\%)  & 46.78 (+105.54\%) \\
        \midrule
        \midrule
        \multirow{6}{*}{C2C}
        & P1 & 09.76 & 39.37 \\
        & P1+P3 & 08.55 (-12.40\%) & 45.28 (+15.01\%) \\
        & P4+P3 & 15.44 (+58.20\%)  & 45.00 (+14.30\%)  \\
        & P5(API)+P3 & 13.37 (+36.99\%)  & 46.17 (+17.27\%) \\
        & P5+P3 & 08.90 (-08.81\%)  & 46.88 (+19.08\%)  \\
        \bottomrule
    \end{tabular}
    \label{tab_prompt_sample}
\end{table}

\subsection{Prompt Design for Text-to-Code Generation}\label{sec_design_t2c}


\ind{Prompt Description.}
As described by Lu et al. \cite{lu2021codexglue}, the T2C generation task takes an NL description as a textual input (e.g., "convert int to String") and expects a correct generation of Java code method, which matches the intent of the description. According to the task description, we naturally present a basic task prompt: \textit{"write a Java method that + \#\{NL\}"} (Table \ref{tab_prompt_type}-P1). To assess the effectiveness of the prompts, we randomly sample 100 instances from training data and ask ChatGPT to generate code given the prompt. We obtain a low generation accuracy of BLEU=5.29 and CodeBLEU=22.76.

\ind{Multi-Step Optimizations.}
With this prompt template (Table \ref{tab_prompt_type}-P1), we asked ChatGPT: \textit{"how to improve the prompt: write a Java method that converts int to string"}. ChatGPT told that by providing more specific details of the method behaviour, programming context, and input/output examples, we can create a more clear and more informative prompt that helps guide the generation of a well-designed Java method. We notice that the programming environment provided in the dataset as described in Section \ref{sec_dataset} can be used as additional context information. To reach the goal, we added a context prompt before the task prompt: \textit{"remember you have a Java class named + '\#\{CN\}', member variables + '\#\{MV\}', member functions + '\#\{MF\}'}  (Table \ref{tab_prompt_type}-P2). In the prompt, the cloze \#{...} will be filled by corresponding information given in the dataset. Note that we tell ChatGPT to remember the class because it will generate the whole class if we do not guide ChatGPT with clear instructions. By adding the context prompt, the accuracy of the samples can be improved with BLEU=10.42 and CodeBLEU=25.05.

After analyzing the ground-truth, we observe that the ground-truth were pre-processed in four aspects: 1) all comments, throws, and method modifiers are removed; 2) the method name is changed to "function"; 3) all the arguments are renamed to "arg0", "arg1", etc.; 4) all the local variables are renamed to "loc0", "loc1", etc. Following these observations, we thus add a processing prompt with a series of instructions after the task prompt: \textit{"remove comments; remove summary; remove throws; remove function modifiers; change method name to "function"; change argument names to "arg0", "arg1"...; change local variable names to "loc0", "loc1"..."}  (Table \ref{tab_prompt_type}-P3). Note that some summaries generated for part of code snippets cannot be removed by the prompt \textit{"remove comments"} but the \textit{"remove summary"}; we used the ellipsis "..." in the prompt instead of "etc.", because ChatGPT cannot do the renaming actions with the command "etc.". The evaluation shows a further improvement with BLEU=13.11 and CodeBLEU=36.00.

By comparing the generated code with the ground-truth, we notice that ChatGPT may generate code with different APIs and settings of exception handling. It is natural to ask ChatGPT to regenerate the code according to its responses and users' specific requirements. To extract the requirement of the APIs and exception handling, we input the ChatGPT with the prompts \textit{"list the used methods with names only in the following Java methods and do not explain: \#\{Code\}"} and \textit{"does the code contain exception handling? + \#\{Code\}"} for the ground-truth, respectively. Afterward, we write scripts to analyze the responses for the requirements of APIs (i.e., name list) and exception handling (i.e., true or false). With these two requirements, we replace the task prompt with a behaviour prompt: \textit{"write a Java method \#\{that calls ...\} with[out] exception handling to \#\{NL\}"}  (Table \ref{tab_prompt_type}-P5). Note that if the API list is empty, we remove the \textit{"\#\{that calls ...\}"}, otherwise we replace \textit{"..."} with the name list. For the \textit{"with[out]"}, we determine whether it is \textit{"with"} or \textit{"without"} according to the actual demand. We find that considering the API requirement, the generation accuracy can be enhanced (BLEU=22.14 and CodeBLEU=44.18). Meanwhile, using the whole behaviour prompt (i.e., API + exception handling), the performance can be further boosted with BLEU=27.48 and CodeBLEU=46.78.

\subsection{Prompt Design for Code-to-Code Generation}\label{sec_design_c2c}

As the C2C generation has a similar process of prompt design as the T2C generation, we mainly show their key differences in this subsection.

\ind{Prompt Description.} 
According to the task description in Section \ref{sec_dataset}, our C2C generation task aims to generate a Java code method according to a given C\# code function. Based on the task requirement, we form the task prompt: \textit{"translate C\# code into Java code: \#\{Code\}"}  (Table \ref{tab_prompt_type}-P1). For the randomly selected 100 samples from training data, the generation performance is BLEU=9.76 and CodeBLEU=39.37.

\ind{Multi-Step Optimizations.}
Comparing with the T2C generation, we can find many differences in the C2C generation task: the C2C dataset does not involve the related class; ChatGPT does not generate comments and throws to the code; the ground-truth are not pre-processed for the method name, modifiers, argument names, and local variable names. However, ChatGPT will generate annotation following the C\# code but the ground-truth removed all the annotations. Therefore, we add a simple processing prompt to the task prompt: \textit{"do not provide annotation"}  (Table \ref{tab_prompt_type}-P3). Moreover, we find that ChatGPT has the ability to understand the markdown syntax in the prompt. Thus, in the task prompt, we change \textit{\#\{Code\}} to \textit{}\textit{'''\#\{Code\}'''} as an updated task prompt  (Table \ref{tab_prompt_type}-P4). Testing on the samples with the processing prompt, the generation accuracy is improved on CodeBLEU (45.28) but not on BLEU (8.55). After updating the code format in the task prompt, we achieve further enhancement with BLEU=15.44 and CodeBLEU=45.00.

Same to T2C generation, we extract the requirement of the API usage and exception handling from the ground-truth code. Subsequently, we added this information to the task prompt as a behaviour prompt: \textit{"translate C\# code into Java code: '''\#\{Code\}''' \#\{that calls ...\} with[out] exception handling"}  (Table \ref{tab_prompt_type}-P5). Experimental results on the samples show that by adding the requirements of API usage (BLEU=13.37 and CodeBLEU=46.17), the generation accuracy will be slightly improved in terms of CodeBLEU. Moreover, using the whole behaviour prompt also showed a reduction in BLEU (8.90) and a minor increase in CodeBLEU (46.88). We observed that ChatGPT can understand translation context and generate good results. But adding more requests may bring uncertainty to the generation. Therefore, this behaviour prompt likely has negative effects on the C2C generation task.

\subsection{Research Questions}\label{sec_design_rqs}

In this study, we propose a method to guide ChatGPT for two code generation tasks. To verify the effectiveness of the method and analyze the associated influential factors, this study investigates the following RQs:


\ind{RQ1: How effective is the designed prompt for ChatGPT?}
As described in Sections \ref{sec_design_t2c} and \ref{sec_design_t2c}, we leveraged CoT strategy \cite{wei2022emergent} to manually augment prompts for two code generation tasks with multi-step optimizations. The first RQ intends to evaluate the effectiveness of the designed prompts on the corresponding testing datasets, and verifies the validity of our design methods.

\ind{RQ2: How does the conciseness request affect ChatGPT?}
In the prompt design, we observed that ChatGPT often generates detailed code, much more complex than the ground truth. Thus, one goal of the multi-step optimizations is to guide ChatGPT to generate concise code with a series of prompts. It is worth investigating whether the generation performance can be further improved by directly requesting ChatGPT for a concise generation.

\ind{RQ3: How does the session setting affect ChatGPT?}
When communicating with ChatGPT, we started one individual session for each prompt. Meanwhile, it is widely known that ChatGPT can learn the session context and generate better responses from the context \cite{ray2023chatgpt,wei2022emergent}. Therefore, this RQ intends to answer whether ChatGPT can generate better code by inputting a session with a number of prompts. 

\ind{RQ4: How does the generation randomness affect ChatGPT?}
It is known that ChatGPT may generate code with slight differences every time for the same prompt \cite{lyu2023translating,reiss2023testing}. To investigate how randomness affects the generation performance, we rerun the guided-ChatGPT multiple times and analyze the stability of the generation performance.


\section{Results}\label{result}
This section presents the experimental settings and results for the four RQs described in Section \ref{sec_design_rqs}. 

\subsection{Effectiveness of the Designed Prompt (RQ1)}\label{rq1}

\ind{Objective.} To assess the effectiveness of the designed prompts, we intend to present some baselines, test them on the testing data of corresponding code generation tasks, and analyze the effectiveness in terms of prediction accuracy. 

\ind{Method.} This study presents three baselines that can generate code by using ChatGPT with three levels of prompts: \textit{1) ChatGPT-task,} we used the task prompts in Table \ref{tab_prompt_type}(P1) as the input of ChatGPT, because they can represent the common chatting conditions with a direct request of code generation. \textit{2) ChatGPT-detail,} we merged the prompts of task, context, processing, and updated task for two tasks in Table \ref{tab_prompt_type}(P1-P4). Note that the T2C generation has no updated task prompt (P4) while the C2C generation contains no context prompt (P2). \textit{3) ChatGPT-behaviour,} based on the baseline ChatGPT-detail, we further updated the task prompt with the behaviour prompt (Table \ref{tab_prompt_type}-P5) to provide more guidance on the code generation. Although Section \ref{sec_design_c2c} showed that P5 has negative effects on the C2C generation task for the sampled data, we still set up this baseline for C2C generation to further confirm the previous observation. The testing data and evaluation metrics were elaborated in Section \ref{study}.

\begin{table}[h]
    \centering
    \caption{Testing three baselines on T2C and C2C generation tasks.}
    \begin{tabular}{l|l|ll}
        \toprule
        \textbf{Task} & \textbf{Model} & \textbf{BLEU} & \textbf{CodeBLEU} \\
        \midrule
        \multirow{3}{*}{T2C}
        &ChatGPT-task & 05.63 & 28.05 \\
        &ChatGPT-detail & 14.09 (+140.27\%) & 39.90 (+42.25\%)\\
        &ChatGPT-behaviour & 21.59 (+283.48\%) & 48.69 (+73.58\%)\\
        \midrule
        \midrule
        \multirow{3}{*}{C2C}
        &ChatGPT-task & 10.61 & 46.12 \\
        &ChatGPT-detail & 15.79 (+48.82\%) & 47.71 (+03.45\%) \\
        &ChatGPT-behaviour & 09.47 (-10.74\%) & 47.38 (+02.32\%)\\
        \bottomrule
    \end{tabular}
    \label{tab_rq1}
\end{table}

\ind{Result.} On the T2C generation task, Table \ref{tab_rq1} shows that ChatGPT-task achieves an generation accuracy of BLEU=5.63 and CodeBLEU=28.05. For the ChatGPT-detail with a number of extended prompts, its generation performance is BLEU=14.09 and CodeBLEU=39.90, outperforming ChatGPT-task by 140.27\% and 42.25\% in terms of BLEU and CodeBLEU respectively. Additionally, the last baseline ChatGPT-behaviour gained a better performance of T2C generation with BLEU=21.59 and CodeBLEU=48.69. We can notice that ChatGPT-behaviour improved the performance of the ChatGPT-task by 283.48\% and 73.58\% in terms of BLEU and CodeBLEU, respectively. These results indicate that our designed prompts associated with the design method can substantially improve the T2C generation for ChatGPT.

In terms of the C2C generation task, Table \ref{tab_rq1} shows that ChatGPT-tasks can obtain better performance (BLEU=10.61 and CodeBLEU=46.12) compared with the T2C generation task. Furthermore, the ChatGPT-detail shows better generation accuracy with BLEU=15.79 and CodeBLEU=47.71, outperforming those of ChatGPT by 48.82\% and 3.45\% respectively. However, we can find that ChatGPT-behaviour shows poorer performance with BLEU=9.47 and CodeBLEU=47.38, increased those of ChatGPT-task by -10.74\% and 2.32\%. Therefore, ChatGPT-detail shows better performance than ChatGPT-behaviour. These results imply that for the C2C task, the designed prompt can also improve the generation performance for ChatGPT.

\answer{1}{Our prompt design method can substantially improve the performance of T2C and C2C generation tasks by guiding ChatGPT with better prompts.}

\subsection{Impacts of Conciseness Request (RQ2)}\label{rq2}

\ind{Objective.} In our prompt design, we guided ChatGPT to remove irrelevant code components. However, we observed that the generated code is still more complex than the ground-truth. Without more information, we cannot design better prompts. Therefore, this RQ intends to investigate whether we can change the task prompt by directly asking ChatGPT to generate a more concise code, instead of manually adding specific instructions.

\ind{Method.} RQ1 showed that ChatGPT-behaviour and ChatGPT-detail are the best baselines for T2C and C2C generation tasks, respectively. To add conciseness requests to their prompts, we found one viable and simple way to add the word "concise" before the generation targets. Specifically, for the T2C generation, the behaviour prompt of ChatGPT-behaviour is changed to \textit{"write a concise Java method ..."} (Table \ref{tab_prompt_type}-P5). Likewise, for the C2C generation task, the task prompt of ChatGPT-detail is updated: \textit{"translate C\# code into concise Java code ..."} (Table \ref{tab_prompt_type}-P4). To evaluate the impacts of the conciseness request, we tested the modified prompts on the testing data.

\begin{table}[h]
    \centering
    \caption{Testing the best baselines in RQ1 with (or without) the conciseness request (C) on T2C and C2C generation tasks.}
    \setlength{\tabcolsep}{6pt}{
    \begin{tabular}{l|p{0.3\linewidth}|ll}
        \toprule
        \textbf{Task} & \textbf{Model} & \textbf{BLEU} & \textbf{CodeBLEU} \\
        \midrule
        \multirow{2}{*}{T2C}
        &ChatGPT-behaviour & 21.59 & 48.69\\
        &ChatGPT-behaviour-C & 26.86 (+24.41\%) & 50.18 (+03.06\%)\\
        \midrule
        \midrule
        \multirow{2}{*}{C2C}
        &ChatGPT-detail & 15.79& 47.71 \\
        &ChatGPT-detail-C & 16.75 (+06.08\%)  & 46.62 (-02.28\%)\\
        \bottomrule
    \end{tabular}}
    \label{tab_rq2}
\end{table}

\ind{Result.} For the T2C generation task, Table \ref{tab_rq2} shows that the conciseness request is helpful (ChatGPT-behaviour-C) with BLEU=26.86 and CodeBLEU=50.18. Compared with the best baseline in RQ1 (ChatGPT-behaviour), these two evaluation metrics are further enhanced by 24.41\% and 3.06\%, respectively. Meanwhile, we notice that by adding the conciseness request to the C2C generation task, the generation performance of ChatGPT-detail-C shows a slight decrease in terms of the CodeBLEU (46.62) compared with the ChatGPT-detail (best baseline in RQ1), although the BLUE score is improved by 6.08\%. These results suggest that the conciseness request is useful for T2C generation but not for C2C generation.

\answer{2}{Adding conciseness request to the prompt can further improve the performance of the T2C generation, but show minor negative effects on the C2C generation.}

\subsection{Impacts of Session Setting (RQ3)}\label{rq3}

\ind{Objective.} By default, we opened an individual session for each prompt and communicated with ChatGPT. In contrast, the communication can be worked with a continuous session that generates responses for a number of prompts. In this way, the ChatGPT can learn from the context and may generate better responses for the code generation tasks. Thus, this RQ aims to analyze the effects of the session setting.

\ind{Method.} During the communication with ChatGPT, we used one session to generate code for a number of prompts. When the number reaches the maximum limit, a further prompt will receive no response. Then, we started another new session. In this way, we can ensure that each session is fully utilized and ChatGPT can better understand the context. We know that the amount of contextual information would be lower for the prior prompts within a session. It is worth nothing to investigate how the session setting affects the code generation performance.

\begin{table}[h]
    \centering
    \caption{Testing the best baseline in RQ2 with (or without) the session setting (S) on T2C and C2C generation tasks.}
    \setlength{\tabcolsep}{5pt}{
    \begin{tabular}{l|p{0.32\linewidth}|ll}
        \toprule
        \textbf{Task} & \textbf{Model} & \textbf{BLEU} & \textbf{CodeBLEU} \\
        \midrule
        \multirow{2}{*}{T2C}
        &ChatGPT-behaviour-C & 26.86 & 50.18\\
        &ChatGPT-behaviour-CS & 29.29 (+09.05\%) & 49.74 (-00.88\%)\\
        \midrule
        \midrule
        \multirow{2}{*}{C2C}
        &ChatGPT-detail & 15.79 & 47.71 \\
        &ChatGPT-detail-S & 16.82 (+06.52\%)  & 48.80 (+02.28\%)\\
        \bottomrule
    \end{tabular}}
    \label{tab_rq3}
\end{table}

\ind{Result.} As shown in Table \ref{tab_rq3}, using the continuous session (ChatGPT-behaviour-CS) showed no overall improvement for the T2C generation with BLEU=29.29 and CodeBLEU=49.74. Compared with the best baseline in RQ2 (ChatGPT-behaviour-C), the BLEU score is improved by 9.05\% but the CodeBLEU is decreased by 0.88\%. On the other hand, the continuous session can present improvement for the C2C generation task (ChatGPT-detail-S) with BLEU=16.82 and CodeBLEU=48.80. Compared with the best baseline ChatGPT-detail in RQ2, the evaluation metrics are further improved by 6.52\% and 2.28\%, respectively. These experimental results indicate that the continuous session is beneficial for the C2C generation task but has no improvement for the T2C generation task. Therefore, the individual session is more suitable for the T2C generation with the designed prompts.

\answer{3}{The continuous session is helpful for the T2C generation task while the individual session is suitable for the C2C generation task.}

\subsection{Impacts of Generation Randomness (RQ4)}\label{rq4}

\ind{Objective.} Commonly, ChatGPT generates responses with slight differences to balance the accuracy and creativity. Thus, the generation randomness may affect the performance of code generation. This RQ investigates how randomness affects the effectiveness of the designed prompts.

\ind{Method.} To reach the goal, we ran the best baselines (i.e., ChatGPT-behaviour-C and ChatGPT-detail-S) in RQ3 five times, respectively. We computed the average (AVG) and standard deviation (STD) of the performance of these multiple generation results. Based on these measurements, we analyzed the generation stability and the effects of the randomness.

\begin{table}[]
    \centering
    \caption{Testing the best baselines in RQ3 on T2C and C2C generation tasks in five rounds (RQ1-R5), where "MIN", "MAX", "AVG", "STD" stand for the minimum, maximum, average and standard deviation of the generation accuracy.}
    \setlength{\tabcolsep}{18pt}{
    \begin{tabular}{c|c|cc}
        \toprule
        \textbf{Task} & \textbf{Model} & \textbf{BLEU} & \textbf{CodeBLEU} \\
        \midrule
        \multirow{9}{*}{T2C}
        & R1 & 26.86 & 50.18\\
        & R2 & 26.85 & 50.07\\
        & R3 & 27.02 & 50.18\\
        & R4 & 26.92 & 50.20\\
        & R5 & 27.00 & 50.17\\\cmidrule{2-4}
        & MIN & 26.86 & 50.07\\
        & MAX & 27.02 & 50.20\\
        & AVG & 26.93 & 50.16\\
        & STD & 00.08 & 00.05 \\
        \midrule
        \midrule
        \multirow{9}{*}{C2C}
        & R1 & 16.82 & 48.80\\
        & R2 & 17.34 & 49.17\\
        & R3 & 17.23 & 48.38\\
        & R4 & 17.32 & 49.15\\
        & R5 & 17.21 & 48.80\\\cmidrule{2-4}
        & MIN & 16.82 & 48.80\\
        & MAX & 17.34 & 49.17\\
        & AVG & 17.18 & 48.86\\
        & STD & 00.21 & 00.33\\
        \bottomrule
    \end{tabular}}
    \label{tab_rq4}
\end{table}

\ind{Result.} From Table \ref{tab_rq4}, we can observe that the five rounds of T2C generation show a stable performance, where BLEU ranges from 26.86 to 27.02 with average=26.93 and STD=0.08; CodeBLEU ranges from 50.07 to 50.20 with average=50.16 and STD=0.05. Meanwhile, the multiple runs of C2C generation also show a stable prediction accuracy, where BLEU ranges from 16.82 to 17.34 with average=17.18 and STD=0.21; CodeBLEU ranges from 48.80 to 49.17 with average=48.86 and STD=0.33. These results indicate that testing ChatGPT with our designed prompts will generate stable responses. We observed that the major reason is that the instructions in the prompts are specific so that the generation randomness is limited. As we observed that the effect of the randomness is negligible, we did not extend this RQ with more rounds of experiments.


\answer{4}{The generation randomness shows little effect on the code generation tasks due to the specific instructions described in the designed prompts.}

\section{Discussion}\label{discuss}
This section provides some qualitative analysis of the designed prompts. Specifically, Section \ref{sec_fine_tune} compares our code generation performance with existing fine-tuned models. Sections \ref{sec_correctness} and \ref{sec_quality} analyze the correctness and quality of the code generated by ChatGPT with our designed prompts.

\subsection{Comparison with Fine-Tuned Models}\label{sec_fine_tune}

\ind{Objective.} Section \ref{result} indicates that the designed prompts can guide ChatGPT to generate substantially better code. This result implies the effectiveness of the \textit{"pre-train, prompt, predict"} paradigm for LLMs as described in Section \ref{sec_pe}. However, many LLMs based on the \textit{"pre-train, fine-tune"} paradigm have been successfully applied in the code generation tasks as exemplified in Section \ref{sec_llm}. Therefore, we would like to compare the performance of these two paradigms. Specifically, we investigate how effective is our method compared with the existing fine-tuned LLMs on code generation tasks.

\ind{Method.} In this study, we tested our method on the widely used dataset CodeXGlue \cite{lu2021codexglue}, which provides a number of benchmark models. We used the experimental results reported by Lu et al. \cite{lu2021codexglue} as comparisons. Moreover, we went through all the related works presented in Section \ref{sec_llm}. We found that Wang et al. \cite{wang2021codet5} also tested their proposed model CodeT5 and baseline model PLBART \cite{ahmad2021unified} on CodeXGlue, so we included these models in our comparisons. We excluded the other reports in the related work because they did not use the CodeXGlue dataset or did not report the CodeBLEU metric following \cite{lu2021codexglue}.

\begin{table*}[]
    \centering
    \caption{Comparisons between the best prompts (ChatGPT-best) and the state-of-the-art fine-tuned LLMs reported in \cite{lu2021codexglue} and \cite{wang2021codet5} on T2C and C2C generation tasks.} 
    \setlength{\tabcolsep}{14pt}{
    \begin{tabular}{l|l|p{0.45\linewidth}|cc}
        \toprule
        \textbf{Task} & \textbf{Model} & \textbf{Description} & \textbf{BLEU} & \textbf{CodeBLEU} \\
        \midrule
        \multirow{22}{*}{T2C}
        &Seq2Seq & An RNN-based sequence to sequence model \cite{sutskever2014sequence}. & 21.31 & 26.39 \\ \cmidrule{2-5}
        & \multirow{2}{*}{Seq2Action+MAML} & A context-aware encoder-decoder mode that leverages model-agnostic meta-learning (MAML) \cite{guo2019coupling}.  & \multirow{2}{*}{24.40} & \multirow{2}{*}{29.46}\\\cmidrule{2-5}
        & \multirow{2}{*}{GPT-2} & Generative Pre-trained Transformer 2 (GPT-2) pre-trained on a very large corpus of English data in a self-supervised fashion \cite{radford2019language}. & \multirow{2}{*}{25.37} & \multirow{2}{*}{29.69} \\\cmidrule{2-5}
        & \multirow{3}{*}{CodeGPT} & A Transformer-based language model that has the same model architecture and training objective of GPT-2 and pre-trained on the programming language (PL) \cite{lu2021codexglue}. & \multirow{3}{*}{28.69} & \multirow{3}{*}{32.71} \\\cmidrule{2-5}
        &CodeGPT-adapted & CodeGPT is continually trained on the code corpus \cite{lu2021codexglue}. & 32.79 & 35.98 \\\cmidrule{2-5}
        & \multirow{2}{*}{PLBART} &  A sequence-to-sequence model capable of performing a broad spectrum of program and language understanding and generation tasks \cite{ahmad2021unified}. &  \multirow{2}{*}{36.69} & \multirow{2}{*}{38.52} \\\cmidrule{2-5}
        & \multirow{3}{*}{CodeT5} & A unified pre-trained encoder-decoder Transformer model that better leverages the code semantics conveyed from the developer-assigned identifiers \cite{wang2021codet5}. &  \multirow{3}{*}{41.48} & \multirow{3}{*}{44.10}\\\cmidrule{2-5}
        & \multirow{2}{*}{\textbf{ChatGPT-best}} & ChatGPT-behaviour-C, the ChatGPT works with context prompt, behaviour prompt, processing prompt, and conciseness request. & \multirow{2}{*}{\textbf{26.86}} & \multirow{2}{*}{\textbf{50.18}}\\
        \midrule
        \midrule
        \multirow{12}{*}{C2C}
        & \multirow{2}{*}{PBSMT} & A traditional phase-based machine translation method that uses statistical models to translate text from one language to another. \cite{zens2002phrase} & \multirow{2}{*}{40.06}  & \multirow{2}{*}{43.48}  \\\cmidrule{2-5}
        & \multirow{2}{*}{\textbf{ChatGPT-best}} & ChatGPT-detail-S, the ChatGPT works with an updated task prompt, processing prompt, and continuous session. & \multirow{2}{*}{\textbf{16.82}} & \multirow{2}{*}{\textbf{48.80}}\\\cmidrule{2-5}
        & \multirow{2}{*}{Transformer} & A sequence-to-sequence encoder-decoder model with self-attention mechanism  \cite{vaswani2017attention}. & \multirow{2}{*}{50.47}  & \multirow{2}{*}{61.59}  \\\cmidrule{2-5}
        & \multirow{2}{*}{CodeBERT} & A bidirectional encoder representations from transformers (BERT) model with pre-trained with six programming languages \cite{feng2020codebert}. & \multirow{2}{*}{72.14}  & \multirow{2}{*}{79.41}  \\\cmidrule{2-5}
        & \multirow{2}{*}{RoBERTa} & It is based on the architecture of the BERT model and is pre-trained on a large corpus of text using a masked language modeling objective \cite{guo2019coupling}. & \multirow{2}{*}{71.99}  & \multirow{2}{*}{80.18} \\\cmidrule{2-5}
        & \multirow{2}{*}{PLBART} &  A sequence-to-sequence model capable of performing a broad spectrum of program and language understanding and generation tasks \cite{ahmad2021unified}. &  \multirow{2}{*}{65.00} & \multirow{2}{*}{85.27}\\
        \bottomrule
    \end{tabular}}
    \label{tab_d1}
\end{table*}

\ind{Result.} Table \ref{tab_d1} illustrated the included fine-tuned LLMs for two code generation tasks, the description and reference of the related LLM, and the reported scores of BLEU and CodeBLEU. We also placed our best prompt settings (ChatGPT-best) in the table. We can find that ChatGPT-best shows the best performance on the T2C generation tasks in terms of CodeBLEU. This result implies that guiding ChatGPT with our designed prompts outperforms the other state-of-the-art fine-tuned LLMs. However, for the C2C generation task, the ChatGPT-best achieved the fifth rank, only outperforming PBSMT \cite{zens2002phrase}. This poorer performance on the C2C task may result from the limited contextual information, so we cannot extend the designed prompt with more specific instructions as shown in Section \ref{rq1}. Moreover, these results can demonstrate the potential capabilities of the \textit{"pre-train, prompt, predict"} paradigm, because the performance could be further improved by providing prompts with more specific instructions or fine-tuning ChatGPT with related training data.

\finding{1}{Guiding ChatGPT with our designed prompt outperforms the state-of-the-art fine-tuned LLMs for the T2C generation, but it shows a poorer rank on the C2C generation due to the limited contextual information expressed in the prompt.}

\subsection{Correctness of Code Generation}\label{sec_correctness}

\ind{Objective.} CodeBLEU (or BLEU) is an effective and widely used metric for automated evaluations. But a generated code with a high score may not be correct. Therefore, we would like to investigate the correctness of the code generated from our best prompt settings.

\ind{Method.} We randomly selected 100 samples from the generated code of each code generation task (T2C and C2C). The correctness is measured by functional equivalency between the generated code and the ground-truth. The relevancy is voted by three authors (the first, second, and fourth), where relevancy is determined when the number of votes is larger than or equal to two. 

\begin{table}[]
    \centering
    \caption{Number of code generated by ChatGPT-best that have functional equivalency with the corresponding ground-truth among the 100 samples.}
    \setlength{\tabcolsep}{30pt}{
    \begin{tabular}{c|cc}
        \toprule
        \textbf{Task} & \textbf{T2C} & \textbf{C2C}\\
        \midrule
        \#Relevancy & 31 & 59\\
        \bottomrule
    \end{tabular}
    \label{tab_d2}}
\end{table}

\ind{Result.} Table \ref{tab_d2} shows that among the 100 samples, the T2C task only generated 31 equivalent code as the ground-truth in functional behaviours. We observed that higher CodeBLEU only indicates correct lexical match, syntactic match, and data flow match, but does not necessarily suggest functional equivalency. Meanwhile, the NL descriptions for T2C generation are usually not specific, and ChatGPT is likely to misunderstand the requirement. A simple example is that for the ground-truth code \textit{"String function()\{return namespaceURI;\}"} which means return the variable namespaceURI, but the NL description provided in the dataset is \textit{"Get the WS-ReliableMessaging namespace to be used for encoding and decoding messages"}. Therefore, for the prompt of T2C generation, the NL description may need to be refined in some ways.

For the C2C generation task, we found 59 generated code snippets that are functionally equivalent to their ground-truth, much better than that for the T2C task. We noticed that the translation from C\# code provides many useful contexts so that ChatGPT can generate the corresponding Java code line by line. However, the T2C generation works only for the code with commonly used APIs. 

\finding{2}{ChatGPT with the best prompt shows better correctness on the C2C task than on the T2C task, because the NL descriptions are not rigorous. } 

\subsection{Quality of Code Generation}\label{sec_quality}

\ind{Objective.} Quality is an important feature of a good generated code, in spite of the correctness of code generation. In this study, we also investigated the quality of the code generated by ChatGPT with our designed prompts. Meanwhile, we also checked the quality of the ground-truth code to compare its quality with the generated code. 

\ind{Method.} To measure the quality of the code generation, we utilized the SonarQube (version 9.8) \cite{sonar2023qube}, a widely used open-source platform for analyzing and tracking the quality of code \cite{lenarduzzi2020sonarqube,marcilio2019static}. It can check three types of quality issues (bug, vulnerability, and code smell) in five severity levels (i.e., blocker, critical, major, minor, and info) \cite{sonar2023qube}. In this study, we measured the quality of code generation by counting the number of code that contain critical or blocker issues. This is because these two severity levels have strong impacts which are commonly considered by developers, where the blocker level has a higher likelihood than the critical.

\begin{table*}[]
    \centering
    \caption{Quality analysis of the code generated by ChatGPT-best and the corresponding ground-truth on T2C and C2C generation tasks. }
    \setlength{\tabcolsep}{15.4pt}{
    \begin{tabular}{c|c|cc|cc|cc|c}
        \toprule
        \multirow{2}{*}{\textbf{Task}} & \multirow{2}{*}{\textbf{Data}} & \multicolumn{2}{c|}{\textbf{\#Bug}} & \multicolumn{2}{c|}{\textbf{\#Vulnerability}} & \multicolumn{2}{c|}{\textbf{\#Code Smell}} & \multirow{2}{*}{\textbf{Total}}\\\cmidrule{3-8}
        && Blocker & Critical & Blocker & Critical & Blocker & Critical & \\
        \midrule
        \multirow{2}{*}{T2C} & Generated Code & 0 & 1 & 0 & 0 & 0& 29 & 30\\
                             & Ground-Truth & 0 & 0  & 0 & 0 & 1 & 34 & 35\\
        \midrule
        \multirow{2}{*}{C2C} & Generated Code & 0 & 0 & 0 & 0 & 56 & 6  & 62\\
                             & Ground-Truth   & 0 & 0 & 0 & 0 & 10 & 7 & 17\\
        \bottomrule
    \end{tabular}}
    \label{tab_d3}
\end{table*}

\ind{Result.} Table \ref{tab_d3} shows that for the T2C generation the code generated by ChatGPT-best shows lightly better quality than the ground-truth. Specifically, SonarQube found one critical bug that reminds us of making sure a local variable is not zero before doing the division. And the other 29 issues belong to the critical code smell that requires further attention. In contrast, the associated ground-truth contain 35 code smells. For the C2C generation, the generated code possess 62 code smells, much higher than the number of the related ground-truth (17 code smells). In these two tasks, the generated code involve no severe bug or vulnerability.

We found that the identified code smells mainly provide six kinds of suggestions: 1) defining a constant instead of duplicating a String; 2) replacing the call of "replaceAll()" by "replace()"; 3) reducing cognitive complexity of a code; 4) adding default case to a switch, not overriding the Object.finalize() method; 5) renaming method to prevent misunderstanding; and 6) using a copy constructor or copy factory instead of "clone" implementation. We believe that it is worthy of addressing the bug and code smells detected by SonarQube, although the number is not large compared to the total count of the dataset. Therefore, ChatGPT may be not able to ensure the quality of the generated code, which require further investigations.

\finding{3}{The code generated by ChatGPT contains no severe bug or vulnerability on the experimented datasets but they contain many code smells, which should be addressed by developers. }

\section{Implication}\label{imp}
Based on our experimental results and discussion, this section provides implications for developers and researchers.

\ind{Tips of Using ChatGPT Prompts for Developers.} Our experimental results showed that guiding ChatGPT with carefully crafted prompts can substantially improve the code generation performance. To make the most of ChatGPT's capabilities, developers are suggested to provide prompts with rich programming context, including relevant classes, member variables, and functions. Additionally, ChatGPT can understand code, so developers can instruct it to preprocess code, such as removing summaries and changing variable names. Enclosing code snippets in ''' could also help ChatGTP better comprehend code blocks with markdown syntax. To generate code with expected APIs or in a concise form, developers can directly request ChatGPT. Chatting with ChatGPT in one continuous session may also be beneficial because it can understand the context and refine its responses. With the chain-of-thoughts feature, developers can optimize their prompts step-by-step by considering ChatGPT's feedback. However, developers should remain cautious about the quality of the generated code, even if the function-level generation appears free of severe bugs or vulnerabilities in our experiments.


\ind{Potential Future Research Directions for Researchers.}
This study demonstrates that providing ChatGPT with improved prompts can enhance code generation performance, outperforming that of state-of-the-art fine-tuned LLMs. Future studies could investigate methods for automatically designing and optimizing prompts for code generation tasks; design a better evaluation metric to automatically assess the correctness of the generated code; and further assess and improve the quality of the code generation. Our study also highlights the potential of LLMs like ChatGPT, utilizing the \textit{"pre-train, prompt, predict"} paradigm and prompt engineering, to impact the field of SE research.


\section{Threats to Validity}\label{threat}
There are some potential threats affecting the validity of our experimental results and conclusions. 

\ind{Limited Tasks and Dataset.} 
This study proposed a method for prompt design on two code generation tasks. The method may not be suitable for other generation tasks, such as code completion \cite{ziegler2022productivity} and test case generation \cite{lukasczyk2023empirical}. For the investigated T2C and C2C generation tasks, we only chose one dataset CodeXGlue \cite{lu2021codexglue} as our study subject, although it is a widely used dataset. On the C2C generation task, we tested the generation from C\# to Java.  It is uncertain whether our experimental results and findings could be extended to other languages (e.g., Python and Go) and the reverse generation (i.e., from Java to C\#). In the near future, we plan to investigate our designed prompts on more programming languages, other datasets, and more types of code generation tasks.

\ind{Manual Prompt Design.} Same as other prompt engineering studies with manual construction, the prompt design and multi-step optimizations are conducted according to human understanding and observations. The knowledge of the designer may affect the performance of the used prompts. Moreover, our design and combination choices were based on the 100 randomly selected samples, and the size and randomness of the sampling may bring in different choices. However, this study aims to explore the viability of prompt design and investigate some related influential factors. Our experimental results demonstrate the effectiveness of the designed prompts. In the future, we would like to investigate prompt engineering with automated construction methods \cite{liu2023pre,qiao2022reasoning}.  

\ind{Limited Testing.}
In our study, we conducted a limited number of testing as presented in Section \ref{result}, such as the combinations of prompt selection and combinations, conciseness request and session settings, and multiple runs for generation randomness. We suppose that more tests may help us further improve the designed prompts and strengthen our conclusions and findings. However, invoking ChatGPT API on the whole dataset requires a high cost. Therefore, we only tested a limited number of choices that are likely to improve the generation performance, so that we can demonstrate the importance of prompt design for ChatGPT.

\ind{Human Evaluation.} 
To assess the correctness of the generated code, we randomly selected 100 samples to analyze the relevancy between the generated code and ground-truth. The relevancy was determined by human evaluations. To mitigate the effects of the subjective bias, the determinations are based on the voting of three authors. Although the number of samples is limited, the correctness analysis of these random samples provided some useful insights for further studies.


\section{Conclusion}\label{conclude}
In this paper, we designed and improved prompts for guiding ChatGPT on two code generation tasks, including text-to-code generation and code-to-code generation. Our experimental results showed the effectiveness of our prompts when asking ChatGPT to generate code on a widely used dataset CodeXGlue. Moreover, we investigated the influential factors for designing prompts on code generation tasks. Besides, we compared the performance of the best prompts with the state-of-the-art fine-tuned LLMs, and assessed the correctness and quality of the code generated by ChatGPT. Based on our findings, we present the potential future research directions. 




\bibliographystyle{IEEEtran}
\bibliography{reference}

\begin{thebibliography}{10}
\providecommand{\url}[1]{#1}
\csname url@samestyle\endcsname
\providecommand{\newblock}{\relax}
\providecommand{\bibinfo}[2]{#2}
\providecommand{\BIBentrySTDinterwordspacing}{\spaceskip=0pt\relax}
\providecommand{\BIBentryALTinterwordstretchfactor}{4}
\providecommand{\BIBentryALTinterwordspacing}{\spaceskip=\fontdimen2\font plus
\BIBentryALTinterwordstretchfactor\fontdimen3\font minus
  \fontdimen4\font\relax}
\providecommand{\BIBforeignlanguage}[2]{{%
\expandafter\ifx\csname l@#1\endcsname\relax
\typeout{** WARNING: IEEEtran.bst: No hyphenation pattern has been}%
\typeout{** loaded for the language `#1'. Using the pattern for}%
\typeout{** the default language instead.}%
\else
\language=\csname l@#1\endcsname
\fi
#2}}
\providecommand{\BIBdecl}{\relax}
\BIBdecl

\bibitem{herrington2003code}
J.~Herrington, \emph{Code generation in action}.\hskip 1em plus 0.5em minus
  0.4em\relax Manning Publications Co., 2003.

\bibitem{poesia2022synchromesh}
G.~Poesia, O.~Polozov, V.~Le, A.~Tiwari, G.~Soares, C.~Meek, and S.~Gulwani,
  ``Synchromesh: Reliable code generation from pre-trained language models,''
  \emph{arXiv preprint arXiv:2201.11227}, 2022.

\bibitem{xu2020incorporating}
F.~F. Xu, Z.~Jiang, P.~Yin, B.~Vasilescu, and G.~Neubig, ``Incorporating
  external knowledge through pre-training for natural language to code
  generation,'' \emph{arXiv preprint arXiv:2004.09015}, 2020.

\bibitem{guo2020graphcodebert}
D.~Guo, S.~Ren, S.~Lu, Z.~Feng, D.~Tang, S.~Liu, L.~Zhou, N.~Duan,
  A.~Svyatkovskiy, S.~Fu \emph{et~al.}, ``Graphcodebert: Pre-training code
  representations with data flow,'' \emph{arXiv preprint arXiv:2009.08366},
  2020.

\bibitem{lu2021codexglue}
S.~Lu, D.~Guo, S.~Ren, J.~Huang, A.~Svyatkovskiy, A.~Blanco, C.~Clement,
  D.~Drain, D.~Jiang, D.~Tang \emph{et~al.}, ``Codexglue: A machine learning
  benchmark dataset for code understanding and generation,'' \emph{arXiv
  preprint arXiv:2102.04664}, 2021.

\bibitem{nijkamp2022codegen}
E.~Nijkamp, B.~Pang, H.~Hayashi, L.~Tu, H.~Wang, Y.~Zhou, S.~Savarese, and
  C.~Xiong, ``Codegen: An open large language model for code with multi-turn
  program synthesis,'' \emph{arXiv preprint arXiv:2203.13474}, 2022.

\bibitem{liu2022commitbart}
S.~Liu, Y.~Li, and Y.~Liu, ``Commitbart: A large pre-trained model for github
  commits,'' \emph{arXiv preprint arXiv:2208.08100}, 2022.

\bibitem{araci2019finbert}
D.~Araci, ``Finbert: Financial sentiment analysis with pre-trained language
  models,'' \emph{arXiv preprint arXiv:1908.10063}, 2019.

\bibitem{zhou2020sentix}
J.~Zhou, J.~Tian, R.~Wang, Y.~Wu, W.~Xiao, and L.~He, ``Sentix: A
  sentiment-aware pre-trained model for cross-domain sentiment analysis,'' in
  \emph{Proceedings of the 28th international conference on computational
  linguistics}, 2020, pp. 568--579.

\bibitem{gunel2020supervised}
B.~Gunel, J.~Du, A.~Conneau, and V.~Stoyanov, ``Supervised contrastive learning
  for pre-trained language model fine-tuning,'' \emph{arXiv preprint
  arXiv:2011.01403}, 2020.

\bibitem{feng2020codebert}
Z.~Feng, D.~Guo, D.~Tang, N.~Duan, X.~Feng, M.~Gong, L.~Shou, B.~Qin, T.~Liu,
  D.~Jiang \emph{et~al.}, ``Codebert: A pre-trained model for programming and
  natural languages,'' \emph{arXiv preprint arXiv:2002.08155}, 2020.

\bibitem{devlin2018bert}
J.~Devlin, M.-W. Chang, K.~Lee, and K.~Toutanova, ``Bert: Pre-training of deep
  bidirectional transformers for language understanding,'' \emph{arXiv preprint
  arXiv:1810.04805}, 2018.

\bibitem{raffel2020exploring}
C.~Raffel, N.~Shazeer, A.~Roberts, K.~Lee, S.~Narang, M.~Matena, Y.~Zhou,
  W.~Li, and P.~J. Liu, ``Exploring the limits of transfer learning with a
  unified text-to-text transformer,'' \emph{The Journal of Machine Learning
  Research}, vol.~21, no.~1, pp. 5485--5551, 2020.

\bibitem{openai2023chatgpt}
OpenAI, ``Chatgpt official blog,'' \url{https://openai.com/blog/chatgpt}, 2023.

\bibitem{khoury2023secure}
R.~Khoury, A.~R. Avila, J.~Brunelle, and B.~M. Camara, ``How secure is code
  generated by chatgpt?'' \emph{arXiv preprint arXiv:2304.09655}, 2023.

\bibitem{ouyang2022training}
L.~Ouyang, J.~Wu, X.~Jiang, D.~Almeida, C.~Wainwright, P.~Mishkin, C.~Zhang,
  S.~Agarwal, K.~Slama, A.~Ray \emph{et~al.}, ``Training language models to
  follow instructions with human feedback,'' \emph{Advances in Neural
  Information Processing Systems}, vol.~35, pp. 27\,730--27\,744, 2022.

\bibitem{kothari2023chatgpt}
A.~Kothari, ``Chatgpt, large language models, and generative ai as future
  augments of surgical cancer care,'' \emph{Annals of Surgical Oncology}, pp.
  1--3, 2023.

\bibitem{liu2022design}
V.~Liu and L.~B. Chilton, ``Design guidelines for prompt engineering
  text-to-image generative models,'' in \emph{Proceedings of the 2022 CHI
  Conference on Human Factors in Computing Systems}, 2022, pp. 1--23.

\bibitem{ren2020codebleu}
S.~Ren, D.~Guo, S.~Lu, L.~Zhou, S.~Liu, D.~Tang, N.~Sundaresan, M.~Zhou,
  A.~Blanco, and S.~Ma, ``Codebleu: a method for automatic evaluation of code
  synthesis,'' \emph{arXiv preprint arXiv:2009.10297}, 2020.

\bibitem{wei2022chain}
J.~Wei, X.~Wang, D.~Schuurmans, M.~Bosma, E.~Chi, Q.~Le, and D.~Zhou, ``Chain
  of thought prompting elicits reasoning in large language models,''
  \emph{arXiv preprint arXiv:2201.11903}, 2022.

\bibitem{salazar2019masked}
J.~Salazar, D.~Liang, T.~Q. Nguyen, and K.~Kirchhoff, ``Masked language model
  scoring,'' \emph{arXiv preprint arXiv:1910.14659}, 2019.

\bibitem{liu2019roberta}
Y.~Liu, M.~Ott, N.~Goyal, J.~Du, M.~Joshi, D.~Chen, O.~Levy, M.~Lewis,
  L.~Zettlemoyer, and V.~Stoyanov, ``Roberta: A robustly optimized bert
  pretraining approach,'' \emph{arXiv preprint arXiv:1907.11692}, 2019.

\bibitem{alokla2022pseudocode}
A.~Alokla, W.~Gad, W.~Nazih, M.~Aref, and A.-b. Salem, ``Pseudocode generation
  from source code using the bart model,'' \emph{Mathematics}, vol.~10, no.~21,
  p. 3967, 2022.

\bibitem{niu2022spt}
C.~Niu, C.~Li, V.~Ng, J.~Ge, L.~Huang, and B.~Luo, ``Spt-code:
  sequence-to-sequence pre-training for learning source code representations,''
  in \emph{Proceedings of the 44th International Conference on Software
  Engineering}, 2022, pp. 2006--2018.

\bibitem{radford2019language}
A.~Radford, J.~Wu, R.~Child, D.~Luan, D.~Amodei, I.~Sutskever \emph{et~al.},
  ``Language models are unsupervised multitask learners,'' \emph{OpenAI blog},
  vol.~1, no.~8, p.~9, 2019.

\bibitem{brown2020language}
T.~Brown, B.~Mann, N.~Ryder, M.~Subbiah, J.~D. Kaplan, P.~Dhariwal,
  A.~Neelakantan, P.~Shyam, G.~Sastry, A.~Askell \emph{et~al.}, ``Language
  models are few-shot learners,'' \emph{Advances in neural information
  processing systems}, vol.~33, pp. 1877--1901, 2020.

\bibitem{vaswani2017attention}
A.~Vaswani, N.~Shazeer, N.~Parmar, J.~Uszkoreit, L.~Jones, A.~N. Gomez,
  {\L}.~Kaiser, and I.~Polosukhin, ``Attention is all you need,''
  \emph{Advances in neural information processing systems}, vol.~30, 2017.

\bibitem{liu2023pre}
P.~Liu, W.~Yuan, J.~Fu, Z.~Jiang, H.~Hayashi, and G.~Neubig, ``Pre-train,
  prompt, and predict: A systematic survey of prompting methods in natural
  language processing,'' \emph{ACM Computing Surveys}, vol.~55, no.~9, pp.
  1--35, 2023.

\bibitem{guo2022unixcoder}
D.~Guo, S.~Lu, N.~Duan, Y.~Wang, M.~Zhou, and J.~Yin, ``Unixcoder: Unified
  cross-modal pre-training for code representation,'' \emph{arXiv preprint
  arXiv:2203.03850}, 2022.

\bibitem{liu2023contrabert}
S.~Liu, B.~Wu, X.~Xie, G.~Meng, and Y.~Liu, ``Contrabert: Enhancing code
  pre-trained models via contrastive learning,'' \emph{arXiv preprint
  arXiv:2301.09072}, 2023.

\bibitem{jain2020contrastive}
P.~Jain, A.~Jain, T.~Zhang, P.~Abbeel, J.~E. Gonzalez, and I.~Stoica,
  ``Contrastive code representation learning,'' \emph{arXiv preprint
  arXiv:2007.04973}, 2020.

\bibitem{mastropaolo2021studying}
A.~Mastropaolo, S.~Scalabrino, N.~Cooper, D.~N. Palacio, D.~Poshyvanyk,
  R.~Oliveto, and G.~Bavota, ``Studying the usage of text-to-text transfer
  transformer to support code-related tasks,'' in \emph{2021 IEEE/ACM 43rd
  International Conference on Software Engineering (ICSE)}.\hskip 1em plus
  0.5em minus 0.4em\relax IEEE, 2021, pp. 336--347.

\bibitem{wang2021codet5}
Y.~Wang, W.~Wang, S.~Joty, and S.~C. Hoi, ``Codet5: Identifier-aware unified
  pre-trained encoder-decoder models for code understanding and generation,''
  \emph{arXiv preprint arXiv:2109.00859}, 2021.

\bibitem{ahmad2021unified}
W.~U. Ahmad, S.~Chakraborty, B.~Ray, and K.-W. Chang, ``Unified pre-training
  for program understanding and generation,'' \emph{arXiv preprint
  arXiv:2103.06333}, 2021.

\bibitem{svyatkovskiy2020intellicode}
A.~Svyatkovskiy, S.~K. Deng, S.~Fu, and N.~Sundaresan, ``Intellicode compose:
  Code generation using transformer,'' in \emph{Proceedings of the 28th ACM
  Joint Meeting on European Software Engineering Conference and Symposium on
  the Foundations of Software Engineering}, 2020, pp. 1433--1443.

\bibitem{husain2019codesearchnet}
H.~Husain, H.-H. Wu, T.~Gazit, M.~Allamanis, and M.~Brockschmidt,
  ``Codesearchnet challenge: Evaluating the state of semantic code search,''
  \emph{arXiv preprint arXiv:1909.09436}, 2019.

\bibitem{chen2021evaluating}
M.~Chen, J.~Tworek, H.~Jun, Q.~Yuan, H.~P. d.~O. Pinto, J.~Kaplan, H.~Edwards,
  Y.~Burda, N.~Joseph, G.~Brockman \emph{et~al.}, ``Evaluating large language
  models trained on code,'' \emph{arXiv preprint arXiv:2107.03374}, 2021.

\bibitem{github2023copilot}
G.~Inc., ``Copilot,'' \url{https://github.com/features/copilot}, 2023.

\bibitem{openai2023gpt4}
OpenAI, ``Gpt-4 technical report,''
  \url{https://doi.org/10.48550/arXiv.2303.08774}, 2023.

\bibitem{rajagopal2021template}
D.~Rajagopal, V.~Khetan, B.~Sacaleanu, A.~Gershman, A.~Fano, and E.~Hovy,
  ``Template filling for controllable commonsense reasoning,'' \emph{arXiv
  preprint arXiv:2111.00539}, 2021.

\bibitem{qiao2022reasoning}
S.~Qiao, Y.~Ou, N.~Zhang, X.~Chen, Y.~Yao, S.~Deng, C.~Tan, F.~Huang, and
  H.~Chen, ``Reasoning with language model prompting: A survey,'' \emph{arXiv
  preprint arXiv:2212.09597}, 2022.

\bibitem{wei2022emergent}
J.~Wei, Y.~Tay, R.~Bommasani, C.~Raffel, B.~Zoph, S.~Borgeaud, D.~Yogatama,
  M.~Bosma, D.~Zhou, D.~Metzler \emph{et~al.}, ``Emergent abilities of large
  language models,'' \emph{arXiv preprint arXiv:2206.07682}, 2022.

\bibitem{zhang2022automatic}
Z.~Zhang, A.~Zhang, M.~Li, and A.~Smola, ``Automatic chain of thought prompting
  in large language models,'' \emph{arXiv preprint arXiv:2210.03493}, 2022.

\bibitem{mishra2022numglue}
S.~Mishra, A.~Mitra, N.~Varshney, B.~Sachdeva, P.~Clark, C.~Baral, and
  A.~Kalyan, ``Numglue: A suite of fundamental yet challenging mathematical
  reasoning tasks,'' \emph{arXiv preprint arXiv:2204.05660}, 2022.

\bibitem{iyer2018mapping}
S.~Iyer, I.~Konstas, A.~Cheung, and L.~Zettlemoyer, ``Mapping language to code
  in programmatic context,'' \emph{arXiv preprint arXiv:1808.09588}, 2018.

\bibitem{papineni2002bleu}
K.~Papineni, S.~Roukos, T.~Ward, and W.-J. Zhu, ``Bleu: a method for automatic
  evaluation of machine translation,'' in \emph{Proceedings of the 40th annual
  meeting of the Association for Computational Linguistics}, 2002, pp.
  311--318.

\bibitem{suzgun2022challenging}
M.~Suzgun, N.~Scales, N.~Sch{\"a}rli, S.~Gehrmann, Y.~Tay, H.~W. Chung,
  A.~Chowdhery, Q.~V. Le, E.~H. Chi, D.~Zhou \emph{et~al.}, ``Challenging
  big-bench tasks and whether chain-of-thought can solve them,'' \emph{arXiv
  preprint arXiv:2210.09261}, 2022.

\bibitem{openai2023api}
OpenAI, ``Chatgpt api,'' \url{https://platform.openai.com/docs/api-reference},
  2023.

\bibitem{ray2023chatgpt}
P.~P. Ray, ``Chatgpt: A comprehensive review on background, applications, key
  challenges, bias, ethics, limitations and future scope,'' \emph{Internet of
  Things and Cyber-Physical Systems}, 2023.

\bibitem{lyu2023translating}
Q.~Lyu, J.~Tan, M.~E. Zapadka, J.~Ponnatapuram, C.~Niu, G.~Wang, and C.~T.
  Whitlow, ``Translating radiology reports into plain language using chatgpt
  and gpt-4 with prompt learning: Promising results, limitations, and
  potential,'' \emph{arXiv preprint arXiv:2303.09038}, 2023.

\bibitem{reiss2023testing}
M.~V. Reiss, ``Testing the reliability of chatgpt for text annotation and
  classification: A cautionary remark,'' \emph{arXiv preprint
  arXiv:2304.11085}, 2023.

\bibitem{sutskever2014sequence}
I.~Sutskever, O.~Vinyals, and Q.~V. Le, ``Sequence to sequence learning with
  neural networks,'' \emph{Advances in neural information processing systems},
  vol.~27, 2014.

\bibitem{guo2019coupling}
D.~Guo, D.~Tang, N.~Duan, M.~Zhou, and J.~Yin, ``Coupling retrieval and
  meta-learning for context-dependent semantic parsing,'' \emph{arXiv preprint
  arXiv:1906.07108}, 2019.

\bibitem{zens2002phrase}
R.~Zens, F.~J. Och, and H.~Ney, ``Phrase-based statistical machine
  translation,'' in \emph{KI 2002: Advances in Artificial Intelligence: 25th
  Annual German Conference on AI, KI 2002 Aachen, Germany, September 16--20,
  2002 Proceedings 25}.\hskip 1em plus 0.5em minus 0.4em\relax Springer, 2002,
  pp. 18--32.

\bibitem{sonar2023qube}
Sonar, ``Sonarqube 9.8,''
  \url{https://docs.sonarqube.org/9.8/user-guide/rules/overview/}, 2023.

\bibitem{lenarduzzi2020sonarqube}
V.~Lenarduzzi, F.~Lomio, H.~Huttunen, and D.~Taibi, ``Are sonarqube rules
  inducing bugs?'' in \emph{2020 IEEE 27th International Conference on Software
  Analysis, Evolution and Reengineering (SANER)}.\hskip 1em plus 0.5em minus
  0.4em\relax IEEE, 2020, pp. 501--511.

\bibitem{marcilio2019static}
D.~Marcilio, R.~Bonif{\'a}cio, E.~Monteiro, E.~Canedo, W.~Luz, and G.~Pinto,
  ``Are static analysis violations really fixed? a closer look at realistic
  usage of sonarqube,'' in \emph{2019 IEEE/ACM 27th International Conference on
  Program Comprehension (ICPC)}.\hskip 1em plus 0.5em minus 0.4em\relax IEEE,
  2019, pp. 209--219.

\bibitem{ziegler2022productivity}
A.~Ziegler, E.~Kalliamvakou, X.~A. Li, A.~Rice, D.~Rifkin, S.~Simister,
  G.~Sittampalam, and E.~Aftandilian, ``Productivity assessment of neural code
  completion,'' in \emph{Proceedings of the 6th ACM SIGPLAN International
  Symposium on Machine Programming}, 2022, pp. 21--29.

\bibitem{lukasczyk2023empirical}
S.~Lukasczyk, F.~Kroi{\ss}, and G.~Fraser, ``An empirical study of automated
  unit test generation for python,'' \emph{Empirical Software Engineering},
  vol.~28, no.~2, p.~36, 2023.

\end{thebibliography}

\end{document}